%% file: main.tex
\documentclass[10pt,conference]{IEEEtran}
\usepackage{cite}
\usepackage{amsmath,amssymb,amsfonts}
\usepackage{algorithmic}
\usepackage{graphicx}
\usepackage{textcomp}
\usepackage{xcolor}
\usepackage{multirow}
\usepackage{float}
\usepackage{subcaption}
\usepackage{hyperref}

\def\BibTeX{{\rm B\kern-.05em{\sc i\kern-.025em b}\kern-.08em
    T\kern-.1667em\lower.7ex\hbox{E}\kern-.125emX}}

\newcommand{\toolname}{ContextModule~}

\begin{document}

%
% The "title" command has an optional parameter, allowing the author to define a "short title" to be used in page headers.
\title{ContextModule: Improving Code Completion via Repository-level Contextual Information}

%
% The "author" command and its associated commands are used to define the authors and their affiliations.
% Of note is the shared affiliation of the first two authors, and the "authornote" and "authornotemark" commands
% used to denote shared contribution to the research.
% @管占明 @刘俊麟 @刘洁瑞 @彭超 @刘德欣 @孙宁远 @江波 MarsCode  @李文超 @刘杰 Jason Liu @朱航
\makeatletter
\newcommand{\linebreakand}{%
  \end{@IEEEauthorhalign}
  \hfill\mbox{}\par
  \mbox{}\hfill\begin{@IEEEauthorhalign}
}
\makeatother

\author{
    \IEEEauthorblockN{Zhanming Guan}
    \IEEEauthorblockA{ByteDance \\
    Shenzhen, China \\
    guanzhanming.steph@bytedance.com}
    \and
    \IEEEauthorblockN{Junlin Liu}
    \IEEEauthorblockA{ByteDance \\
    Shenzhen, China \\
    liujunlin.jerry@bytedance.com}
    \and
    \IEEEauthorblockN{Jierui Liu}
    \IEEEauthorblockA{ByteDance \\
    Hangzhou, China \\
    liujierui.0723@bytedance.com}

    \linebreakand

    \IEEEauthorblockN{Chao Peng\textsuperscript{*}}
    \IEEEauthorblockA{ByteDance \\
    Beijing, China \\
    pengchao.x@bytedance.com}
    \and
    \IEEEauthorblockN{Dexin Liu}
    \IEEEauthorblockA{ByteDance \\
    Hangzhou, China \\
    liudexin.ldx@bytedance.com}
    \and
    \IEEEauthorblockN{Ningyuan Sun}
    \IEEEauthorblockA{ByteDance \\
    Shenzhen, China \\
    sunningyuan@bytedance.com}

    \linebreakand

    \IEEEauthorblockN{Bo Jiang}
    \IEEEauthorblockA{ByteDance \\
    Shenzhen, China \\
    jiangbo.jacob@bytedance.com}
    \and
    \IEEEauthorblockN{Wenchao Li}
    \IEEEauthorblockA{ByteDance \\
    Shenzhen, China \\
    liwenchao.24@bytedance.com}
    \and
    \IEEEauthorblockN{Jie Liu}
    \IEEEauthorblockA{ByteDance \\
    Hangzhou, China \\
    liujie.jl@bytedance.com}
    \and
    \IEEEauthorblockN{Hang Zhu}
    \IEEEauthorblockA{ByteDance \\
    Hangzhou, China \\
    zhuhang.leon@bytedance.com}
}

% \author{
%     \IEEEauthorblockN{Zhanming Guan, Junlin Liu, Jierui Liu, Chao Peng, Dexin Liu, \\ Ningyuan Sun, Bo Jiang, Wenchao Li, Jierui Liu, Hang Zhu}
%     \IEEEauthorblockA{ByteDance \\
%     China \\
%     guanzhanming.steph@bytedance.com}
% }

\maketitle

\begingroup\renewcommand\thefootnote{*}
\footnotetext{Corresponding author.}
\endgroup

\begin{abstract}
\input{abstract}
\end{abstract}


\section{Introduction}
\label{sec:introduction}
\input{introduction}

\section{Background}
\label{sec:background}
\input{background}

\section{Approach}
\label{sec:approach}
\input{approach}

\section{Evaluation}
\label{sec:experiment}
\input{experiment}

% \section{Results}
% \label{sec:result}
% \input{result}

\section{Related Work}
\label{sec:related}
\input{related}

\section{Conclusion and Future Work}
\label{sec:conclusion}
\input{conclusion}

\bibliographystyle{IEEEtran}
\bibliography{references}

\end{document}

%% file: abstract.tex
Large Language Models (LLMs) have demonstrated impressive capabilities in code completion tasks, where they assist developers by predicting and generating new code in real-time.
However, existing LLM-based code completion systems primarily rely on the immediate context of the file being edited, often missing valuable repository-level information, user behaviour and edit history that could improve suggestion accuracy.
Additionally, challenges such as efficiently retrieving relevant code snippets from large repositories, incorporating user behavior, and balancing accuracy with low-latency requirements in production environments remain unresolved.
In this paper, we propose ContextModule, a framework designed to enhance LLM-based code completion by retrieving and integrating three types of contextual information from the repository: user behavior-based code, similar code snippets, and critical symbol definitions.
By capturing user interactions across files and leveraging repository-wide static analysis, ContextModule improves the relevance and precision of generated code.
We implement performance optimizations, such as index caching, to ensure the system meets the latency constraints of real-world coding environments.
Experimental results and industrial practise demonstrate that ContextModule significantly improves code completion accuracy and user acceptance rates.

%% file: introduction.tex
Large Language Models (LLMs) have shown remarkable performance in code-related tasks in the industry~\cite{copilot,Codeium,marscode}, driven by the extensive knowledge of programming languages and frameworks acquired during pre-training.
Among these code-related tasks, code completion has emerged as one of the most impactful applications, where LLMs assist developers by predicting and generating the code they need in real time.
This capability significantly boosts development efficiency by reducing the amount of manual coding required and has given rise to notable code assistants in the form of IDE (Integrated Development Environment) plugins, such as GitHub Copilot, Cursor and MarsCode, etc.

While much of the current focus of code completion is on the immediate file that a developer is working on, leveraging repository-level contextual information provides additional opportunities to improve code completion.
This context includes core function definitions, configuration files, and code snippets from elsewhere in the repository that align with the developer's current logic.
By incorporating this broader context, LLMs can generate more accurate and context-aware code, ultimately increasing the acceptance rate of suggestions and further enhancing coding efficiency.

Although both the academia and the industry have explored how to leverage repository-level information to help LLMs predict code, several practical challenges remain under-explored, particularly in real-world production environments. These challenges include:

\begin{itemize}
    \item \textbf{Challenge 1. Limited Contextual Understanding}: Current systems rely mainly on the file being edited, lacking broader repository-level context such as relevant function definitions and similar code snippets.
    \item \textbf{Challenge 2. User Behavior and Intent Recognition}: Current approaches do not effectively incorporate developers' cross-file browsing and editing behavior, missing valuable intent information.
	\item \textbf{Challenge 3. Context Retrieval in Large Repositories with Low Latency}: Retrieving contextual information from large repositories and ensuring accurate suggestions while meeting strict latency requirements in real-world production environments is a significant challenge.
\end{itemize}

To address these challenges, we propose \toolname, a framework designed to retrieve and utilise repository-level contextual information dynamically during the code completion process.
\toolname retrieves three key types of context: (1) \textbf{user behavior-based code}, (2) \textbf{similar code snippets}, and (3) \textbf{critical symbol definitions}.
These retrievals are concatenated with the current file content to form a prompt, which is then passed to the LLM for code generation. The system leverages user browsing and editing behavior, repository-wide code similarities, and key symbol definitions, enabling the model to generate more accurate and context-aware completions.

\begin{itemize}
    \item \textbf{User Behavior Code}: \toolname tracks the developer's recent interactions across different files, such as browsing or editing history.
    This information provides insights into the developer's intent, even when the content is not directly related to the current code.
    By analyzing the user's cursor history, \toolname retrieves relevant code snippets from recently accessed files, offering valuable contextual cues that improve the LLM's predictions.
	\item \textbf{Similar Code}: Developers often reuse patterns or structures that already exist in the repository.
    \toolname employs a text similarity-based retrieval method to identify snippets that share functional similarities with the user's current coding task.
    To achieve this efficiently, we implemented a caching mechanism that reduces retrieval latency, ensuring the system meets the strict performance requirements of production environments.
	\item \textbf{CKG-based Symbol Definition}: Retrieving accurate symbol definitions for key methods, classes, and structures is crucial for reducing the hallucination problem in LLMs.
    \toolname integrates a Code Knowledge Graph (CKG), built through static code analysis, to quickly access key symbol information during the completion process.
    This enables the LLM to generate code that better aligns with the developer's intentions by referencing precise symbol contexts such as function signatures and class structures.
\end{itemize}

% The architecture of \toolname is illustrated in Figure~\ref{fig:total}.
% \toolname first tracks the developer's interaction history, including the files they have browsed and edited, capturing click timestamps and file paths.
% This information reflects the developer's coding intent and helps refine the accuracy of code completions.
% Next, based on the developer's current position in the code, \toolname searches the repository for similar code snippets, which often contain functional logic that can aid completion.
% Lastly, by leveraging static code analysis, \toolname extracts key symbol definitions, such as methods, classes, and data structures, to further reduce the likelihood of incorrect or irrelevant completions.
% To meet the low-latency demands of production environments, we implement an index caching mechanism and introduce a code knowledge graph, to optimise retrieval speed while maintaining a balance between performance and accuracy.

To ensure the system meets the low-latency demands of real-world coding environments, we implemented several optimization techniques, including an index caching mechanism and an incremental parsing system.
These ensure that repository-wide searches remain efficient, maintaining a balance between accuracy and performance.

We evaluated \toolname on multiple datasets, tailored to different types of contextual information, and observed significant improvements in evaluation metrics.
Moreover, the industrial deployment of \toolname into our company's internal code completion system has also demonstrated notable increases in user acceptance rates, which further validates the effectiveness of \toolname.

In summary, we make the following contributions in this paper:

\begin{enumerate}
    \item We present the design of \toolname, a framework that enhances code completion by retrieving relevant repository-level context, including user behavior-based code, similar code, and symbol definitions.
    \item We addressed practical performance challenges, offering optimisation solutions that balance accuracy and low-latency requirements in production environments.
    \item \toolname has been deployed within the company's in-house code completion system, where it has achieved significant improvements in user acceptance rates in both offline evaluations and real-world usage.
\end{enumerate}

%% file: background.tex
\subsection{LLMs and Code Completion}

LLMs, such as GPT-4, DeepSeek and Doubao, have emerged as powerful tools for a wide range of natural language processing (NLP) tasks, including text generation, question answering, and text translation.
These models are pre-trained on vast amounts of data, enabling them to learn complex language patterns, including those found in programming languages.
By leveraging the same principles that allow them to understand and generate natural language, LLMs can also generate high-quality code, making them valuable assets in software development.

One of the most prominent applications of LLMs in the software engineering domain is code completion, where the model assists developers by predicting and generating code snippets in real-time as they write.
Code completion tools, powered by LLMs, provide suggestions based on the developer's current context, such as the surrounding code, function names, or documentation.
This functionality significantly enhances productivity by reducing the amount of manual coding, helping developers write faster and more efficiently.

\subsection{Challenges and Motivation}

Despite the remarkable advancements in LLM-based code completion, several challenges remain when deploying these models in real-world environments:

\subsubsection{Limited Contextual Understanding}
While relying on the content of the file the user is working on can generate useful suggestions, it often lacks the broader repository-level context that can be crucial for more complex tasks.
For instance, relevant function definitions, configuration files, or similar code snippets elsewhere in the repository may not be considered, limiting the model's ability to generate the most accurate or context-aware code.

\subsubsection{User Behavior and Intent Recognition}
Developers frequently browse and edit code across multiple files while working on a particular task.
This cross-file behavior contains implicit information about the developer's intent that can be valuable for code generation.
However, current code completion systems do not effectively incorporate this user behavior, resulting in suggestions that may not fully align with the developer's immediate goals.
Integrating user behavior into the completion process presents both a technical and design challenge.

\subsubsection{Efficient Retrieval of Repository-Level Information with Low Latency}
In large repositories, retrieving the necessary contextual information can be time-consuming and computationally expensive.
For example, developers working on a large project may need relevant code snippets, key function signatures, or symbol definitions from different parts of the repository to guide their current task.
However, real-world deployment of LLM-powered tools introduces strict latency requirements, especially in code completion, where developers expect suggestions to appear almost instantly.
The challenge is to balance accuracy, ensuring relevant context is considered, with performance, keeping response times within acceptable limits.

\subsubsection*{Motivation}
Given these challenges, the motivation for this paper is to explore how repository-level contextual information can be effectively integrated into LLM-based code completion.
We aim to address the limitations of current systems by proposing \toolname, a framework designed to retrieve and utilize broader repository context, user behavior, and critical symbol definitions during the code completion process.
By tackling issues related to context retrieval, user behavior tracking, and performance optimization, this work seeks to improve the accuracy, relevance, and efficiency of LLM-based code completion in production environments.

%% file: approach.tex
\begin{figure*}[htbp]
    \centering
    \includegraphics[width=2\columnwidth]{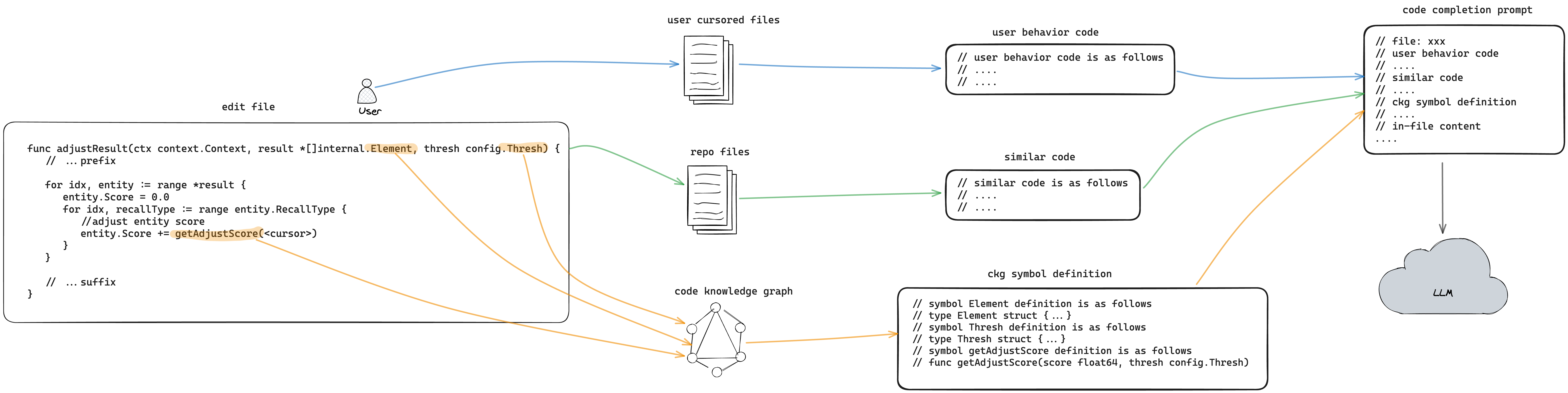}
    \caption{FrameWork of \toolname}
    \label{fig:total}
\end{figure*}

In this section, we present the design and workflow of \toolname based on the observations of the coding process of our developers, as illustrated in Figure~\ref{fig:total}.
During code completion, \toolname retrieves three types of contextual information from the repository where the user is working on.
These retrievals are concatenated with the current file content to form a prompt, which is then passed to the LLM for code generation.
In the rest of this section, we describe in detail the retrieval strategy for each type of contextual information.

\subsection{User Behavior Code}
\subsubsection{Observation}
During development, users frequently engage in cross-file coding and browsing activities.
These actions provide critical insights into the user's intent, even if the code they browse or edit in other files doesn't closely resemble the code at their current position.
For example, users may:

\begin{itemize}
    \item Browse other files to reference the implementation of some logic;
    \item Browse other files to reference certain function parameter definitions;
    \item Implement helper methods elsewhere in the repository.
\end{itemize}

These interactions carry valuable information, enhancing the LLM's ability to generate code more aligned with the user's goals.

\subsubsection{Retrieval Strategy}

\begin{figure*}[htbp]
    \centering
    \includegraphics[width=0.7\textwidth]{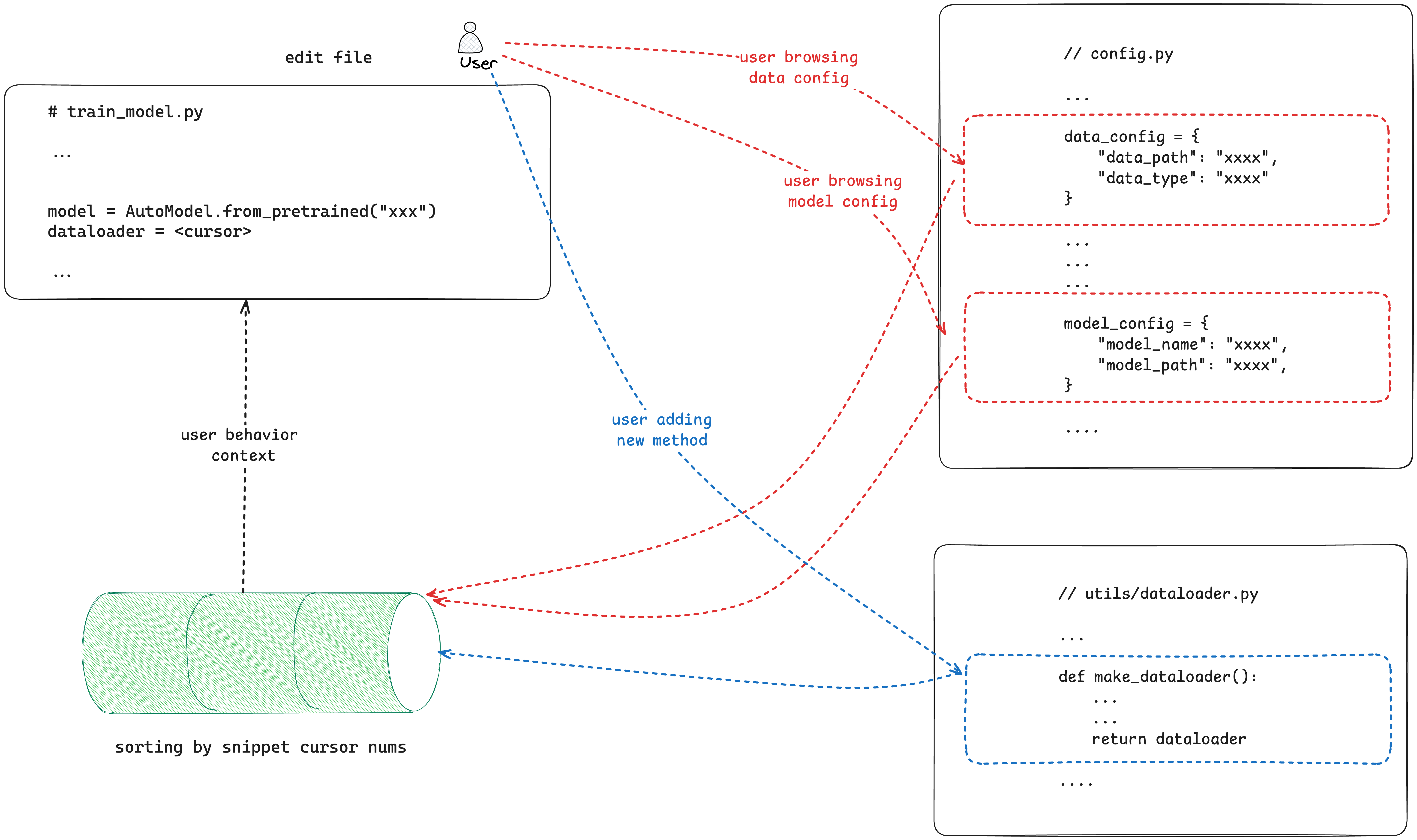}
    \caption{User Behavior Code Retrieval Process}
    \label{fig:user}
\end{figure*}

Each time a user browses or edits code, their cursor activity (clicks, file path, and line number) is recorded.
We analyze this cursor history to extract code snippets related to the user's recent behavior.
Specifically, we retrieve the $N$ most recently browsed files (default is set to $5$) and divide them into snippets using a sliding window.
We then count cursor clicks in each snippet and select the top $K$ snippets with the highest interaction counts as the retrieval results.
The process is illustrated in Figure~\ref{fig:user}.

    % \vspace{-10pt}

\subsection{Similar Code}
\subsubsection{Observation}

During the coding process, users often refer to the existing code in the repository to achieve similar functionality. For example:
\begin{itemize}
    \item When adding a new API, there could be a lot of reusable or similar code for logging and database read/write operations;
    \item When implementing model training logic, there might exist some reusable code for data loading and model training.
\end{itemize}

Such similar snippets, closely related to the user's current task, serve as valuable references for code completion.

\subsubsection{Retrieval Strategy}

We retrieve similar code snippets based on text similarity.
First, all files in the repository are divided into snippets using a sliding window.
Text features are then extracted for each snippet and stored in an indexed knowledge base.
Upon triggering code completion, we extract features from the code near the cursor and use it to search for similar snippets in the knowledge base.
The retrieved snippets are sorted by similarity scores.

After comparing token-based and embedding-based retrieval methods, we chose a token-based approach due to its superior performance and lower implementation cost.
Tokens are extracted using regular expressions, with further splitting of snake and camel case, followed by stop-word removal.
Snippets are ranked by Jaccard similarity scores.

\subsubsection{Caching Strategy}
\begin{figure}[htbp]
    \centering
    \includegraphics[width=\columnwidth]{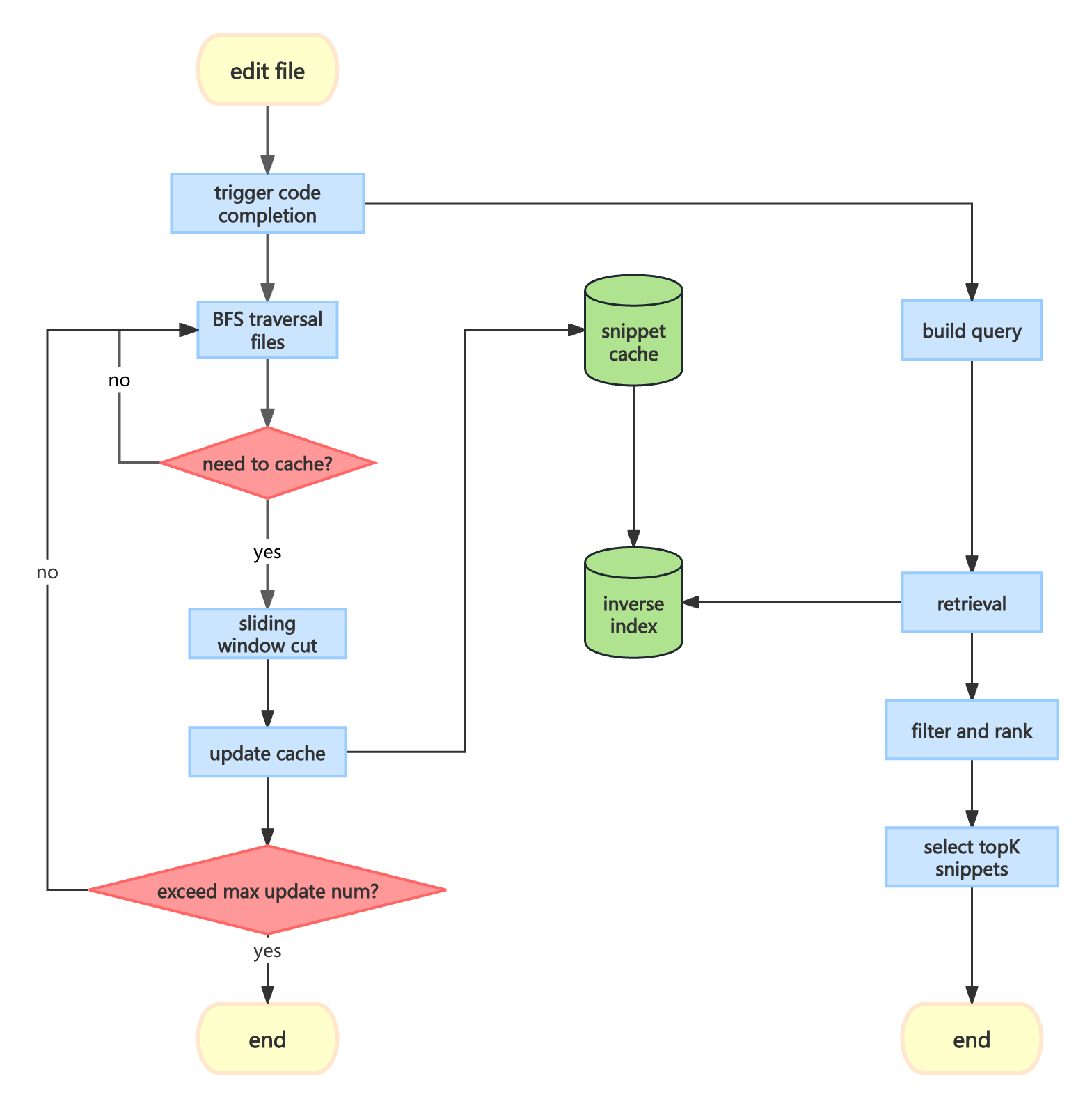}
    \caption{Cache Strategy for Similar Code Retrieval}
    \label{fig:cache}
\end{figure}

In production environments, it is common to have very large repositories containing tens of thousands to hundreds of thousands of files.
Additionally, code completion has strict latency requirements, typically requiring a response time of a few hundred milliseconds.
Therefore, it is not practical to index the entire repository and perform a full search and continuous update of the indexed knowledge base.
To address this issue, we designed a caching scheme, as shown in figure~\ref{fig:cache}.
The caching process includes:

\paragraph{File Sorting Strategy}
We designed a file sorting strategy to determine the priority of different files during the construction and update of the index cache.
For different levels of directories, we use a BFS strategy (current directory - subdirectory - parent directory).
For files in the same directory level, we determine their priority by comparing the longest common prefix of the target file name and the current file name.

\paragraph{Index Cache}
For a given ordered file sequence, we build an index for each file and store it in a cache queue, with a set queue limit (default value is $3000$).
We maintain a set of inverted index systems for all snippets in the cache, which can retrieve all indexed snippets containing any given token.

\paragraph{Index Update}
When a user creates a new file, the index update mechanism is triggered.
Specifically, this mechanism builds an index for the sorted files and updates it in the cache.
To avoid performance degradation, we set a limit on the number of cache updates per cycle, which practically cannot exceed 1/10 of the queue capacity limit.

    % \vspace{-10pt}

With this strategy, we achieve retrieval latency below 80ms in production environments.

\subsection{CKG-based Symbol Definition}
\subsubsection{Observation}
Accurate symbol definitions, such as methods, classes, and structures, are essential for code completion. These definitions reduce LLM hallucinations and improve the accuracy of generated code. Examples include:

\begin{itemize}
    \item Retrieving method signatures for parameter completion;
    \item Accessing class structures for initializing objects;
    \item Collecting class methods and member variables to predict the next steps in coding.
\end{itemize}

Initially, we used the Language Server Protocol (LSP)~\cite{lsp} to retrieve symbol definitions.
However, it posed performance and accuracy challenges, leading to issues with latency and redundant information:

\paragraph{Performance delay}: The LSP typically takes several hundred milliseconds to retrieve symbol definitions, which leads to slow code completion.
\paragraph{Caching issues}: Cached symbol definitions are often outdated or inaccurate at the time of completion, limiting the effectiveness of caching.
\paragraph{Multiple hops}: The LSP may require multiple steps to retrieve information, causing redundancy and performance degradation.
\paragraph{Accuracy limitations}: The LSP sometimes fails to accurately capture the required information at jump positions, necessitating complex post-processing, which increases latency.
\paragraph{Redundant information}: Retrieving all symbols near the cursor results in excessive data, which can obscure important symbols and reduce the effectiveness of code completion models.

\begin{figure*}[htbp]
    \centering
    \includegraphics[width=0.8\textwidth]{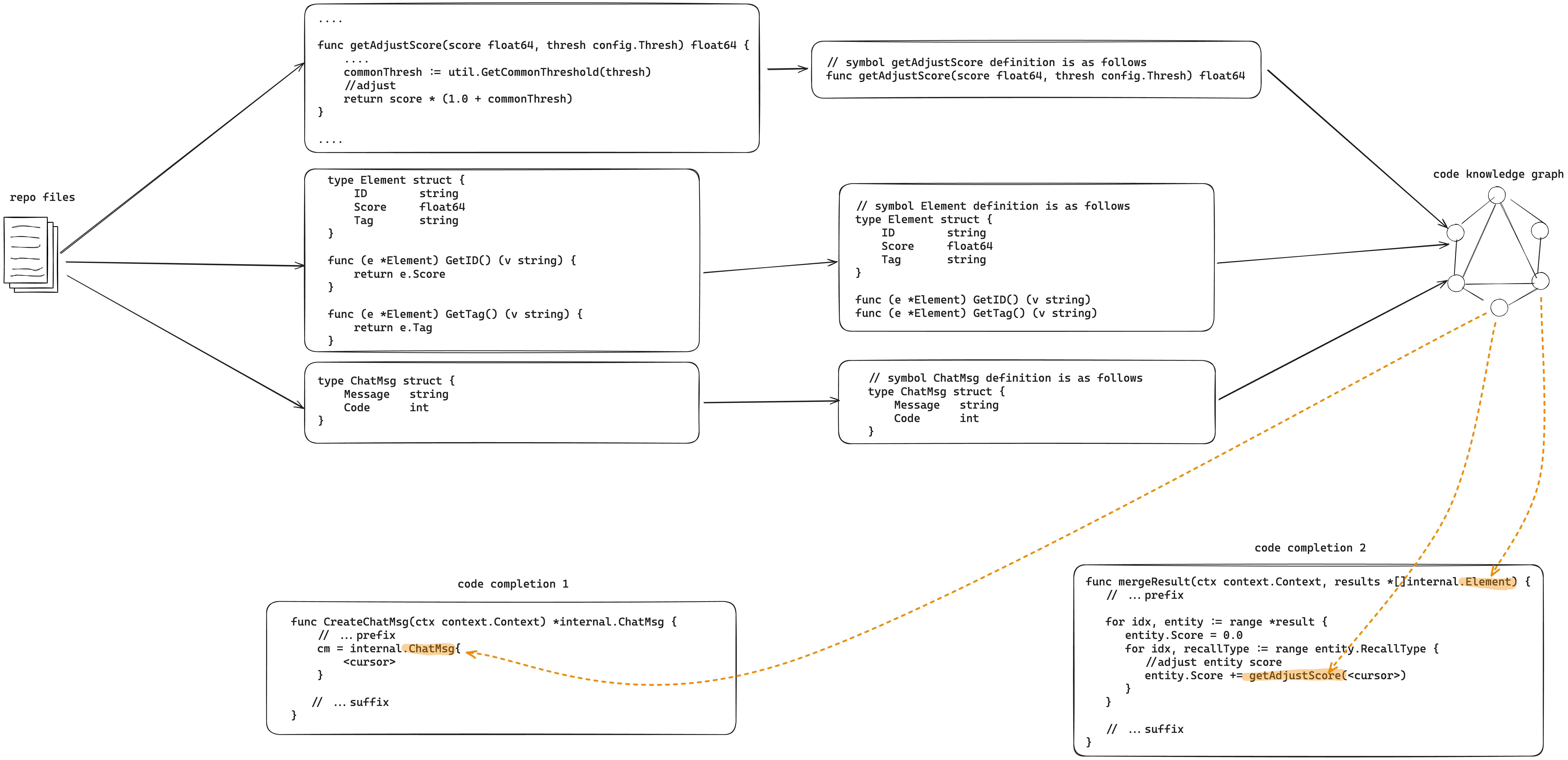}
    \caption{CKG-based Symbol Retrieval Process}
    \label{fig:ckg}
\end{figure*}
    % \vspace{-10pt}

\subsubsection{CKG-based Retrieval Strategy}
Based on the above issues, we replaced LSP with our company's proprietary tool named CKG (code knowledge graph).
CKG is based on code static analysis techniques and constructs a code knowledge graph by analyzing key symbol definitions and dependencies in the repository, enabling us to quickly search for key information in the repository.

Currently, we have implemented a CKG-based retrieval solution and defined two types of key contexts:

\begin{itemize}
    \item \textbf{Function}: We use the function name as the symbol name and the function signature as the symbol context.
    \item \textbf{Struct}: We use the structure name as the symbol name and the structure definition and method signatures as the symbol context.
\end{itemize}

The overall construction process and format of the CKG is shown in Figure~\ref{fig:ckg}.
During the initialization phase, CKG parses all functions and structures in the repository and stores them in the code knowledge graph.
During the developer's coding process, we continuously perform incremental parsing on the user's coding files.
We have pre-defined several key scenarios, and when the completion position is triggered in these scenarios, the corresponding symbol information will be collected.
Key scenarios information include:

\begin{itemize}
    \item \textbf{Function Call}: During the completion of function parameters, we obtain the function signature based on the function name and place it in the prompt.
    \item \textbf{Struct Initialization}: During structure initialization, we obtain the structure information based on the structure name and place it in the prompt. 
    \item \textbf{Function Body}: When code completion is triggered at any position in the function body, we collect the symbol information in the function parameters and place it in the prompt.
\end{itemize}

With the code knowledge graph and incremental parsing capabilities provided by CKG, we can implement single retrieval within 50ms.

%% file: experiment.tex
In this section, we provide an empirical and practical evaluation of ContextModule, analyzing its effectiveness across different types of retrieved context and their impact on code completion in the production environment.

\subsection{Experiment Setup}
\subsubsection{Model Selection and Inference Parameters}
We chose \textbf{DeepSeek-Coder-6.7b-Base}~\cite{guo2024deepseekcoderlargelanguagemodel} as the experimental model for this study.
Since our focus is on assessing the impact of different retrieval strategies within ContextModule, we did not compare our results against other models.

During inference, we employed greedy search to reduce the effect of randomness, setting the maximum input length to $4000$ tokens and the maximum generation length to $100$ tokens. These settings are sufficient for the majority of the test cases.

\subsubsection{Prompt Structure}

We used the \textbf{fill-in-the-middle}~\cite{bavarian2022efficienttraininglanguagemodels} format to construct the prompts, combining the retrieved context with the prefix and suffix from the current file.
Fill-in-the-middle is a prompt format where the model is given both the beginning (prefix) and the end (suffix) of a code snippet, and the task is to generate the missing code in between.
This format allows the model to leverage both preceding and following context when predicting the missing code, improving the accuracy of its completions.
The retrieved context is added in the form of comments to minimize interference with the model's understanding, ensuring that the additional information guides the model without altering the code structure.

\subsubsection{Metrics}

We evaluated the model's performance using two metrics: \textbf{edit similarity} and \textbf{soft exact match}.

\begin{itemize}
  \item \textbf{Edit Similarity: } This metric calculates the ratio of the Levenshtein edit distance between two strings to their total length, providing a measure of how similar the generated code is to the ground truth.
  \item \textbf{Soft Exact Match: } This is an improvement over traditional exact match, which only checks if the generated output matches the label exactly. Soft exact match considers a result correct as long as the retrieved context helps the model generate the correct answer, even if the output doesn't stop precisely at the right point, reducing false negatives caused by unnecessary trailing content.
\end{itemize}

\subsection{Evaluation Dataset Construction}

We constructed three datasets to evaluate different types of context: User Behavior Code, Similar Code, and CKG-based Symbol Definition. For user behavior code and similar code, evaluation datasets were built for Go, Python, and TypeScript with 1500 samples per language. The symbol definition dataset was built only for Go. We used tree-sitter~\footnote{\url{https://tree-sitter.github.io/tree-sitter}} for syntax parsing during the dataset construction.

% The three context strategies in ContextModule are implemented and applied at different stages, and their retrieval methods and application scenarios are also different.
% Therefore, for each type of context, we construct different datasets to conduct experiments and verify their effectiveness.
% In the following, we introduce the details of several datasets.
% For the datasets of user behavior code and similar code, we construct them for three languages: Go, Python and TypeScript, while for the dataset of symbol definition, we only construct it for Go language.
% For each language in each dataset, we construct 1500 data samples.
% The syntax parsing involved in the dataset construction process is completed using tree-sitter.

\subsubsection{User Behavior Code}

To construct this dataset, we relied on real user editing behavior from within our company.
We collected user editing data during code completion and recorded the final code at the completion point, which can either be the code generated by the model or the code written by the user.
Data where the in-file code was stable (i.e. the context of code completion is not changed frequently) and the generated code was correct were selected through a process combining rule-based filtering and manual annotation.
These instances served as the basis for evaluating the impact of user behavior code on completion.

\subsubsection{Similar Code}
We select repositories with active code contributions within the company and construct a dataset of similar code by "digging holes" (i.e., removing parts of the code in the file and serving as the point for the model to complete code) in the file contents. Specifically,  two strategies were employed to remove code:
\begin{itemize}
    \item For each language, we pre-define a series of syntax scenarios as shown in Table~\ref{table:similar code scenario}. We dig holes based on syntax scenarios to construct test data;
    \item We randomly select the starting position for digging holes and use the end of the current syntax block as the end of the hole;
\end{itemize}

\begin{table*}[htbp!]
    \centering
    \caption{Similar Code Scenario}
    \begin{tabular}{ccp{10cm}}   % 定义了三列，每列的对齐方式分别为左对齐、居中对齐、右对齐
    \hline                       % 添加一条横线
    \textbf{Scenario} & \textbf{Language} & \textbf{Description} \\   % 插入表格的列名
    \hline
    function  definition & Python, Go, TypeScript & Keep function signature and predict function body. \\ 
    if-block & Python, Go, TypeScript & Keep if condition and predict if block. \\ 
    function call & Python, Go, TypeScript & Keep call function name and predict the arguments. \\ 
    expression & Python, Go, TypeScript & Predict an expression statement. \\ 
    return statement & Python, Go, TypeScript & Predict return content. \\ 
    arrow function & TypeScript & Predict the arrow function body. \\ 
    JSX element & TypeScript & Predict the content between JSX elements. \\ 
    JSX element attributes & TypeScript & Predict the attributes of JSX elements.\\ 
    \hline
    \end{tabular}
    \label{table:similar code scenario}
\end{table*}

We dig holes and sample within the entire repository range according to the above strategies and mix the test data of the two strategies in a $1:1$ ratio.
We also add extra processes to improve the validity of the data, including ensuring that the predicted content does not appear in the in-file prefix and suffix to avoid additional hints to the model.

\subsubsection{CKG-based Symbol Definition}
To construct the CKG-based symbol dataset, we also select our internal repositories with frequent code contributions, similar to the dataset construction process for similar code.
We targeted specific scenarios such as function calls, struct initialization, and function body predictions.
Similar to the similar code dataset, we applied sampling and filtering strategies to maintain data quality.

\begin{itemize}
  \item \textbf{Function Call}:
We collect function call scenarios, keep the function name, and use the function parameters as the label.
  \item \textbf{Struct Initialization}:
We collect struct initialization scenarios, keep the struct name, and use the initialization code as the label.
  \item \textbf{Function Definition}:
We keep the function signature and use the function body as the label. For the position where the parameter is used in the function body, we keep the parameter and use the content after the parameter as the label.
\end{itemize}

\section{Experimental Results and Online Performance}

\subsection{User Behavior Code as Context}
\begin{table*}[]
\caption{User Behavior Results}
\centering
\begin{tabular}{ccccccc}
    \hline
    \multirow{2}*{} & \multicolumn{2}{c}{Python} & \multicolumn{2}{c}{Go} & \multicolumn{2}{c}{TypeScript} \\
    ~ & {SEM} & {Edit Similarity} & {SEM} & {Edit Similarity} & {SEM} & {Edit Similarity} \\
    \hline
      Without User Behavior Code & 22.5 & 50 & 22.35 & 53.9 & 19.2 & 48.4 \\
      With User Behavior Code & \textbf{24.15} & \textbf{51.9} & \textbf{24.15} & \textbf{55.7} & \textbf{21.75} & \textbf{50.8} \\
    \hline
\end{tabular}
\label{table:user behavior code result}
\end{table*}

We compared the code completion performance of the subject model with and without user behavior code, where we used the top 2 user behavior codes as context.
As shown in the Table~\ref{table:user behavior code result}, after adding user behavior code, the soft exact match (SEM) metric is improved by 1.65\% to 2.55\%, and the edit distance score is improved by 1.8\% to 2.4\%.
This indicates that user behavior code contains information about the user's coding intent, which helps improve code completion effectiveness.
Considering that this improvement comes from real user completion data collected online, it is consistent with online metrics and further validates the important role of user behavior code.

\subsection{Similar Code as Context}
\begin{table*}[]
\caption{Similar Code Results}
\centering
\begin{tabular}{cccccccc}
    \hline
    \multirow{2}*{Strategy} & \multirow{2}*{Max Length of Token} & \multicolumn{2}{c}{Python} & \multicolumn{2}{c}{Go} & \multicolumn{2}{c}{TypeScript} \\
    ~ & & {SEM} & {Edit Similarity} & {SEM} & {Edit Similarity} & {SEM} & {Edit Similarity} \\
    \hline
      Without Context & 4k & 49.77 & 72 & 44.93 & 67.2 & 27.26 & 58.13 \\
      With Context \& Origin & 4k & 53.65 & 76.02 & 53.6 & 72.67 & 37.4 & 63.3 \\
      With Context \& Cut & 4k & \textbf{56.55} & \textbf{77.55} & 54.5 & 73.1 & \textbf{39.13} & \textbf{64.2} \\
      With context \& embedding & 4k & 55.91 & 74.75 & \textbf{55.73} & \textbf{74.02} & 37.35 & 62.63 \\
      With Context \& Cut & 8k & \textbf{59.1} & \textbf{80.58} & \textbf{56.3} & \textbf{75.57} & \textbf{45.3} & \textbf{68.65} \\
    \hline
\end{tabular}
\label{table:similar code result}
\end{table*}

The results of similar code as context are shown in Table~\ref{table:similar code result}.
We compare and analyze the model performance from two aspects: different \textbf{text feature extraction strategies} and \textbf{maximum input length}.

\subsubsection{Text Feature Extraction Strategy}
We compare different text feature extraction methods, including two token extraction strategies and one embedding retrieval strategy:

\begin{itemize}
  \item \textbf{Origin Token}: Using regular expressions to extract tokens such as variables, methods, and parameters from all code as text features.
  \item \textbf{Cut Token}: For origin tokens named in camel and snake formats, we further split them based on underscores and capitalization. Then, we filter the entire token using stop words, including keywords from different code languages and natural language stop words provided by nltk~\cite{nltk}.
  \item \textbf{Embedding}: Using codet5p~\cite{wang2023codet5} model to embed the code and obtain a text vector containing semantic information.
\end{itemize}

For token-based retrieval, we use Jaccard~\cite{bag2019efficient} score to calculate similarity and set $0.1$ as the filtering threshold. For embedding-based retrieval, we use cosine similarity and set $0.7$ as the filtering threshold. We set the window size of code snippets to $30$ lines. After retrieving from the repository, we sorte and filtered the results based on similarity scores and remove snippets with overlapping lines. Finally, we select the top $2$ similar code snippets as the context.

As shown in the table, retrieving similar code from the repository can greatly improve code completion performance, validating the effectiveness of similar code. The cut strategy can further improve the performance by 1\%-3\% on top of the original strategy. This is mainly because there are many tokens in camel and snake formats in the code, and splitting them can obtain more fine-grained semantic information, helping to retrieve more similar code snippets.
For example, the tokens \textit{getUserCreditInfo} and \textit{getUserBasicInfo} have high semantic similarity, which can be captured by the cut strategy but not the origin strategy.
The performance of embedding-based retrieval is similar to that of the cut strategy, better in Python but worse in Go and TypeScript. This may be because we directly use sliding windows to slice the code when constructing code snippets, which destroy the semantic structure of the code and make it difficult for the embedding model to perform well. Considering the high implementation cost of the embedding model, we have not explored it further for now.

\subsubsection{Model Max Length}
We compare the impact of similar code snippets on the model's maximum length at $4k$/$8k$. For a maximum length of $8k$, we increase the window size to 60 lines and select the top $4$ code snippets to concatenate into the prompt. The results show that increasing the maximum length to 8k further improves the model's performance by about 2\%-6\%. This is because a larger search range provides more opportunities to retrieve similar code that can help with code completion. This also further validates the effectiveness of similar code.

\subsection{CKG-based Symbol Definition}
\begin{table*}[]
\caption{CKG-based Symbol Definition Results}
\centering
\begin{tabular}{ccccccc}
    \hline
    \multirow{2}*{} & \multicolumn{2}{c}{Function Call} & \multicolumn{2}{c}{Struct Initialization} & \multicolumn{2}{c}{Function Declaration} \\
    ~ & {soft exact match} & {edit similarity} & {soft exact match} & {edit similarity} & {soft exact match} & {edit similarity} \\
    \hline
      w.o context & 7.58 & 51.08 & 6.17 & 49.19 & 24.1 & 59.45 \\
    \hline
      w. ckg symbol & \textbf{9.34} & \textbf{54.32} & \textbf{17.41} & \textbf{64.19} & \textbf{25.75} & \textbf{60.75} \\
    \hline
\end{tabular}
\label{table:ckg result}
\end{table*}
We compare the effectiveness of adding symbol information in three scenarios: \textbf{function definition}, \textbf{function call}, and \textbf{struct initialization}.
The results shown in Table~\ref{table:ckg result} indicate that adding symbol information significantly improves the performance in all three scenarios, fully validating the effectiveness of ckg-based symbol definition.
In the function definition scenario, using all input parameters as context leads to a 1\%-2\% improvement in relevant metrics.
This indicates that providing input parameter information to LLM can help it better understand the logic of the function to be implemented and generate content more accurately.
Similarly, there is a significant improvement in the function call and struct initialization scenarios, with the improvement in struct initialization exceeding 10\%.
This is mainly because adding context solves the model's hallucination problem and helps it generate parameters accurately.
Function call parameters are relatively simple and can obtain some information from in-file content, while struct initialization is more difficult. This is why there is a significant difference in the improvement between the two scenarios.

\subsection{Context Fusion and Online Performance}
We did not conduct offline experiments on context fusion (combing all context retrieval strategies), we report the performance gain in production environments in this section.
Specifically, we concatenate the context in order of symbol, similar code, and user behavior code (each context having a maximum length), and combine it with the prefix and suffix of the current file to form a completion prompt.
We adapt a rule-based fusion strategy instead of other ranking algorithms such as BM25~\cite{bm25} and embedding similarity to sort retrieval fragments.
This is mainly because the three types of context provide features for code completion from different dimensions, while most ranking algorithms can only compare semantic similarity and are difficult to provide truly effective sorting.
For example, a code fragment that a user has browsed may have a very low similarity to the current file code, but it contains the key configuration that the user needs now.
For the above reasons, we manually set priorities and length limits to complete the fusion of context.

In terms of online performance, we implemented three strategies in sequence of user behavior code, similar code and symbol definition, achieving relative accept rate improvements of 7.1\%, 6.3\%, and 4.9\%, respectively.
It should be noted that the three strategies were implemented in order, and there might be overlapping effects on the results, so the above indicators could not fully represent the effectiveness of each strategy.
We did not conduct separate or overall ablation experiments on these strategies in the production environment, but according to our estimates, the overall improvement could achieve a relative increase of more than 15\%.

\subsection{Case Study}

We select representative cases from real user completion to demonstrate the positive impact of different contextual information.
Considering privacy and space issues, we blurred the code and only displayed the key parts.

\paragraph{CKG Case 1}

\begin{figure*}[!htbp]
  \centering
  \includegraphics[width=0.7\textwidth]{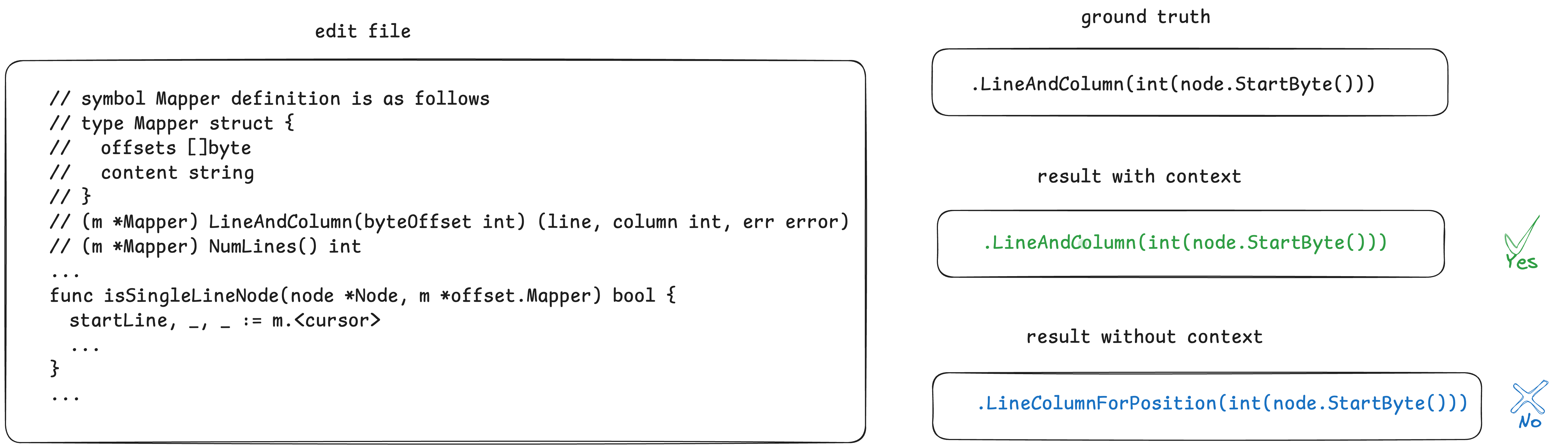}
  \caption{CKG Case 1}
  \label{fig:ckg-case-1}
\end{figure*}

As shown in Figure~\ref{fig:ckg-case-1}, the user needs to call \textit{LineAndColumn} method in \textit{Mapper} to obtain variable \textit{startLine} when implementing \textit{isSingleLine} method. By adding the struct definition and method signature of \textit{Mapper}, the model accurately generates \textit{LineAndColumn} method. Without this context, the model generates \textit{LineColumnForPosition} method, which does not exist.

\paragraph{CKG Case 2}

\begin{figure*}[!htbp]
  \centering
  \includegraphics[width=0.7\textwidth]{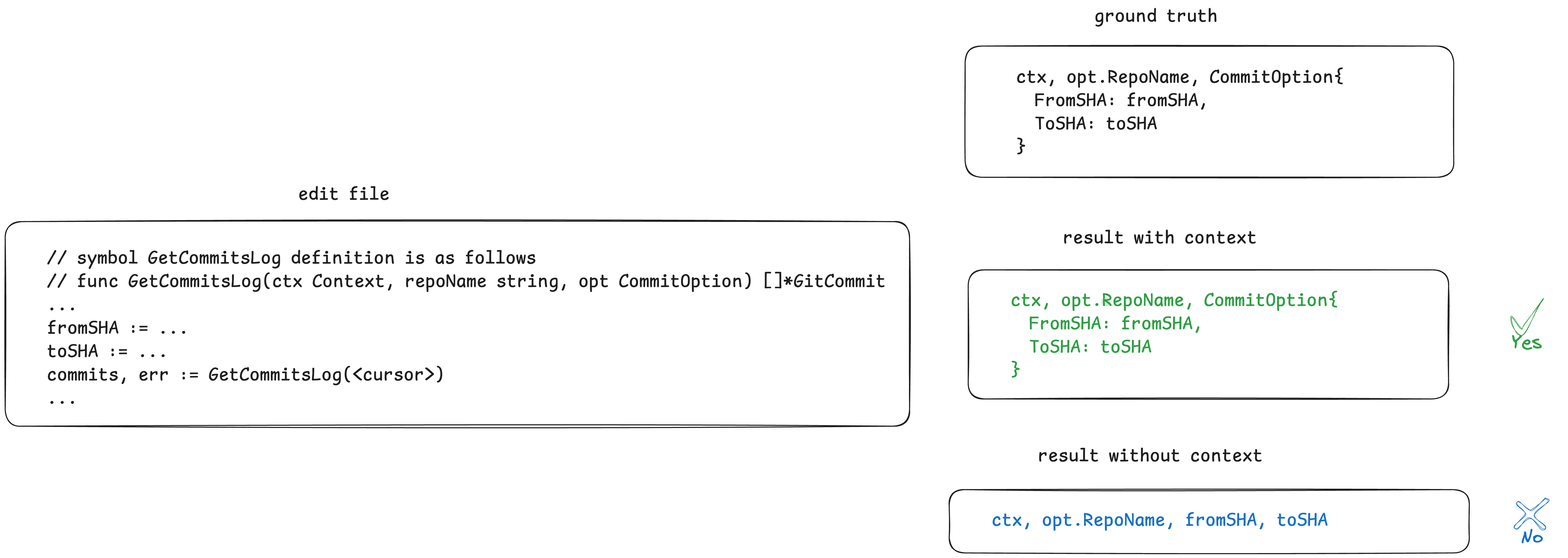}
  \caption{CKG Case 2}
  \label{fig:ckg-case-2}
\end{figure*}

In the completion shown in Figure~\ref{fig:ckg-case-2}, the user needs to call \textit{GetCommitsLog} method to obtain commit information based on \textit{fromSHA} and \textit{toSHA}. By adding the signature information of \textit{GetCommitsLog} method, the model knows that it needs to construct \textit{CommitOption} based on \textit{fromSHA} and \textit{toSHA} as a method parameter. Without this context, the model directly passes \textit{fromSHA} and \textit{toSHA} to the method, which causes the method call failure.

\paragraph{CKG Case 3}

\begin{figure*}[!htbp]
  \centering
  \includegraphics[width=0.7\textwidth]{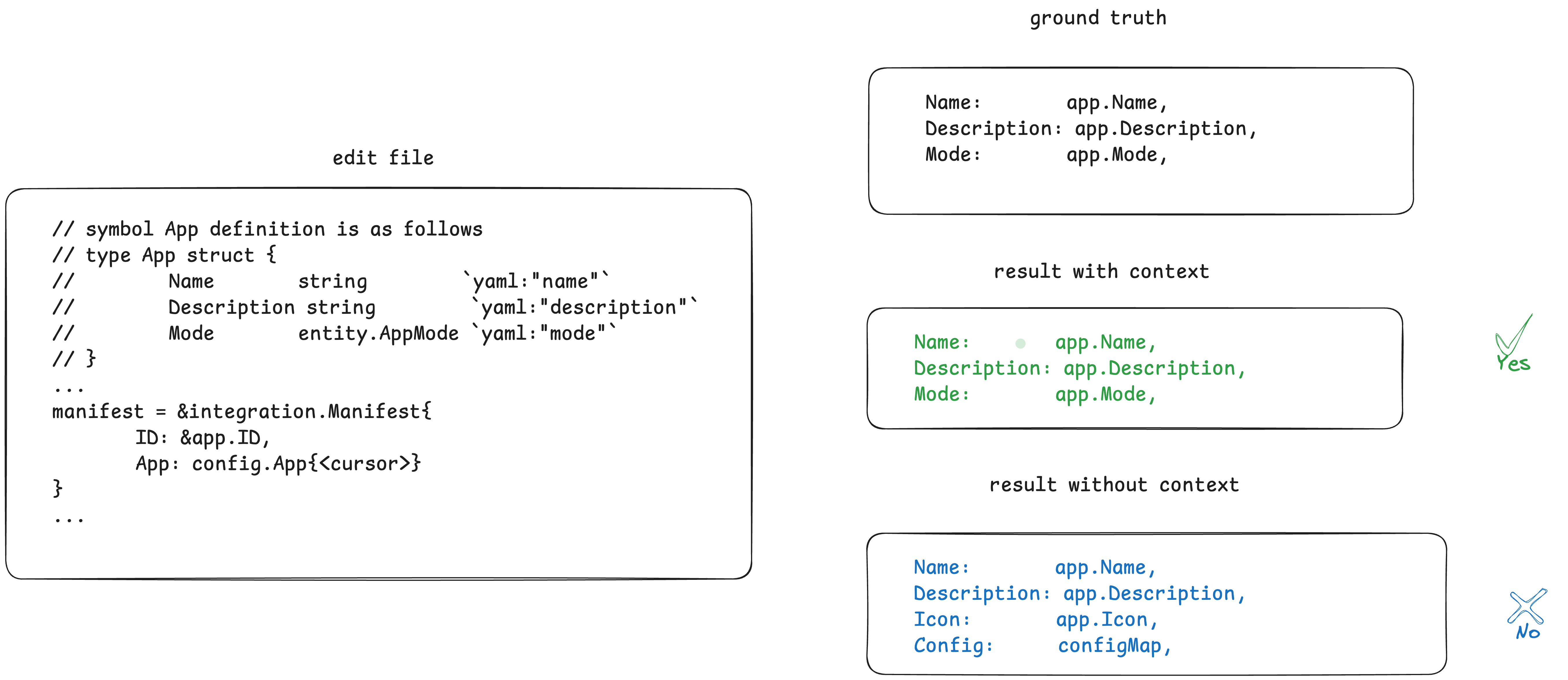}
  \caption{CKG Case 3}
  \label{fig:ckg-case-3}
\end{figure*}

As illustrated in Figure~\ref{fig:ckg-case-3}, the user needs to initialize \textit{App} struct. By adding the struct definition of \textit{App}, the model correctly assigns values to several properties including \textit{Name}, \textit{Description}, and \textit{Mode}. Without the context, the model incorrectly generates \textit{Icon} and \textit{Config} properties.

\paragraph{User Behavior Case 1}

\begin{figure*}[!htbp]
  \centering
  \includegraphics[width=0.7\textwidth]{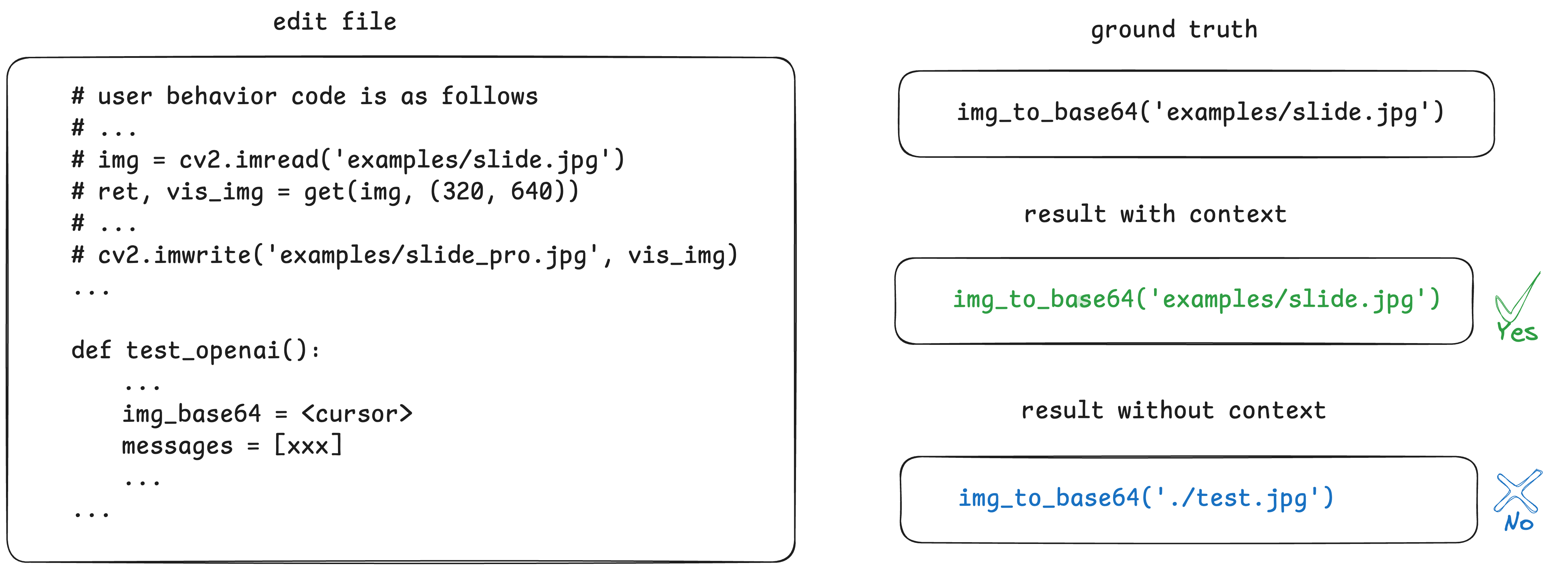}
  \caption{User Behavior Case 1}
  \label{fig:user-case-1}
\end{figure*}

In Figure~\ref{fig:user-case-1}, the user needs to call \textit{img\_to\_base64} method to convert the specified image to the base64 type for downstream tasks. By searching the user behavior code, we found that the user had just loaded and processed the image under \textit{examples/slide.jpg}. The model used this information well to generate accurate path information for the user. Without this context, the model can only simulate a path for recommendation.

\paragraph{User Behavior Case 2}

\begin{figure*}[!htbp]
  \centering
  \includegraphics[width=0.65\textwidth]{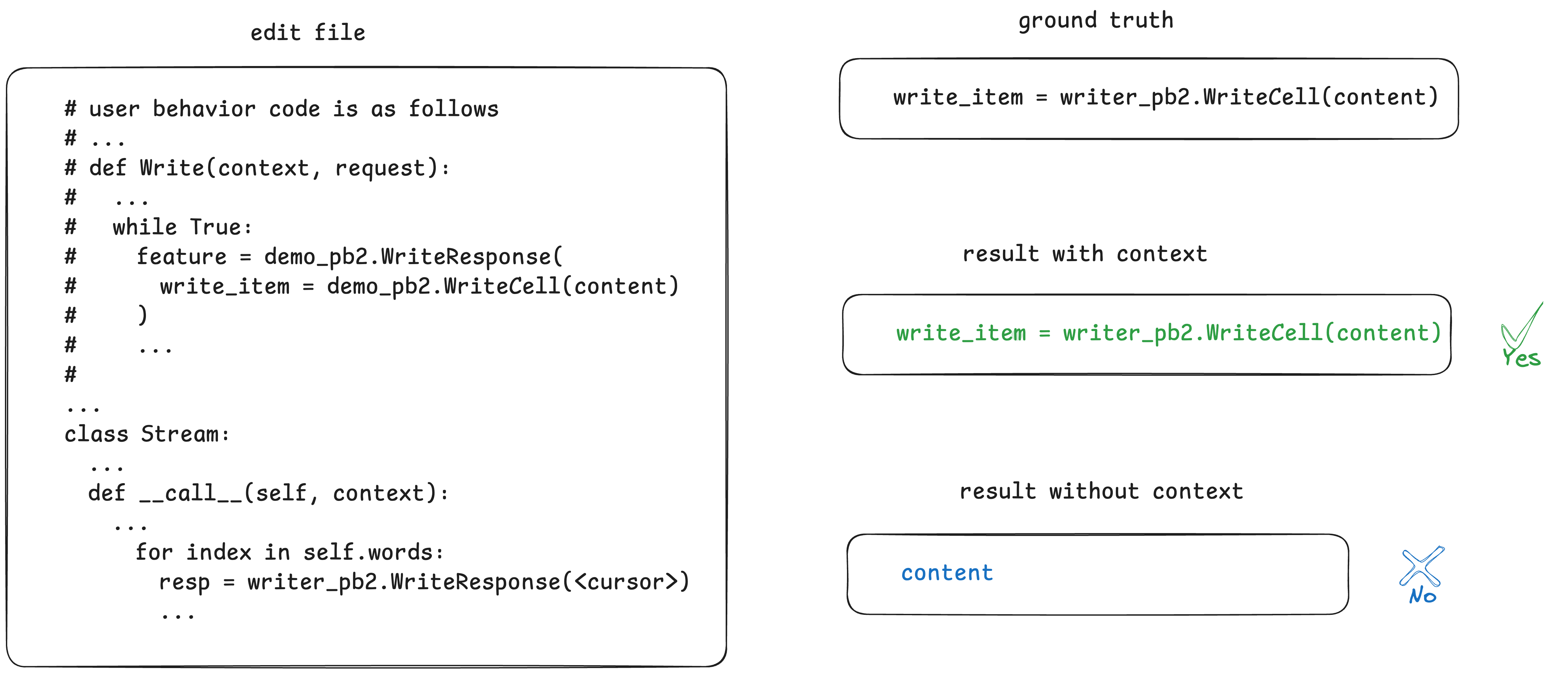}
  \caption{User Behavior Case 2}
  \label{fig:user-case-2}
\end{figure*}

As shown in Figure~\ref{fig:user-case-2}, the user needs to rely on \textit{Writer\_pb2.WriteCell} method to process variable content as the \textit{write\_item} parameter when calling \textit{writer\_pb2.WriteResponse} method. By searching the user behavior code, we found that there is a similar \textit{WriteResponse} method call under \textit{demo\_pb2} that provides a good example for the model to generate the answer accurately. Without this context, the model does not know the existence of \textit{WriteCell} method and generates the parameter incorrectly.

\paragraph{Similar Code Case}

\begin{figure*}[!htbp]
  \centering
  \includegraphics[width=0.65\textwidth]{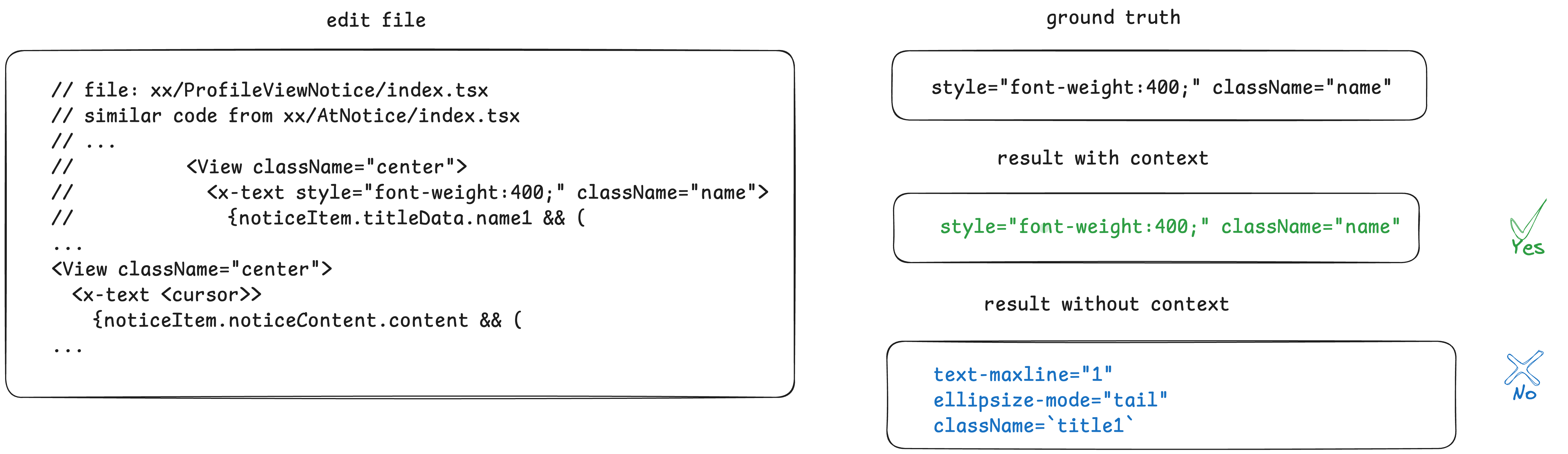}
  \caption{Similar Code Case}
  \label{fig:similar-case-1}
\end{figure*}

In completion shown in Figure~\ref{fig:similar-case-1}, the user needs to initialize the \textit{style} and \textit{className} parameters for \textit{x-text} component in \textit{ProfileViewNotice} page. By searching the repository for similar code, we found similar code logic in \textit{AtNotice} page, which helps the model accurately generate the two parameters and values. Without the context, the model generates \textit{text}, \textit{ellipsize-mode}, and \textit{className} parameters incorrectly.

%% file: related.tex
In recent years, the application of LLMs for code completion has garnered significant attention in both academia and industry~\cite{husein2024large, hou2023large, guo2022unixcoder, li2023structured, li2023large}.
Code completion has traditionally been viewed as a token prediction task~\cite{izadi2022codefill}, but it has evolved to encompass line and block completion, where entire lines or blocks of code are predicted~\cite{svyatkovskiy2020intellicode,du2023classeval,copilot}.
This evolution has been accompanied by the development of increasingly sophisticated models that can utilize both left-context (preceding the cursor) and, in some cases, bi-directional context to predict missing code segments~\cite{fried2022incoder}.

Several empirical studies have highlighted the challenges and gaps in current code completion tools~\cite{izadi2024language}.
Proksch et al.~\cite{proksch2016evaluating} and Hellendoorn et al.~\cite{hellendoorn2019code} emphasized that many evaluations of code completion models are performed using synthetic benchmarks, which do not accurately capture real-world usage patterns.
In live programming environments, models often struggle with the complexity and fluidity of developer interactions, resulting in decreased performance~\cite{aye2021learning}.
These studies further revealed the shortcomings of existing models in real-world contexts, noting that out-of-vocabulary tokens and incomplete context understanding are common causes of failure in auto-completions.

To satisfy these practical needs, researchers have put effort on training code-specific LLMs~\cite{roziere2023code, zhu2024deepseek, lozhkov2024starcoder, guo2023longcoder}, optimizing model architecture~\cite{chen2024bridge, zhu2023improving} and utilising supervised fine-tuning (SFT) to improve model's performance~\cite{liu2020multi, ciniselli2021empirical} on code completion.
In addition to model training, some works explore prompt strategies to help models better understand the context and user requirements~\cite{li2023structured,li2023large, zhang2023planning, mu2024clarifygpt}.
For instance, RepoCoder~\cite{zhang2023repocoder} leverages off-the-shelf context retrieval tools to gather useful information within the repository and enhance the context for LLMs.
RepoHyper~\cite{phan2024repohyper} designs a new repository representation and implements a expand-and-refine retrieval strategy to gather related repository information.

However, existing approaches lack fine-grained strategies for incorporating user behavior or retrieving symbols across multiple files in real-time.
Our work aims to address these limitations by combining user behaviour code, similar code and CKG-based symbol definition retrieval with an index caching mechanism to meet the stringent latency requirements of real-world coding environments without compromising accuracy.

%% file: conclusion.tex
In this paper, we presented ContextModule, a framework designed to enhance code completion by retrieving three types of contextual information: user behavior code, similar code snippets, and CKG-based symbol definitions.
ContextModule significantly improves code prediction accuracy while meeting the low-latency demands of production environments through the implementation of an index caching mechanism and a code knowledge graph.
Both offline experiments and real-world deployment within our internal code completion system demonstrated substantial improvements in key performance metrics and user acceptance rate.

In future work, we plan to extend our work in several directions.
First, we aim to expand the Code Knowledge Graph (CKG) to support additional programming languages and leverage dataflow parsing techniques to retrieve variable types and related definitions, further enhancing context retrieval.
Second, we intend to build a remote knowledge base to incorporate methods and components from open-source repositories, recognizing the growing reliance on external libraries in modern software development.
Finally, we will explore more advanced methods for capturing user intent by analyzing the content of code edits in greater detail, complementing the existing use of cursor history to provide even more precise code completions.